\documentclass[a4paper,11pt]{article}
\usepackage{graphicx}
\usepackage{fullpage}
\usepackage{framed}

\usepackage{mystyle} 
\graphicspath{{./figures/}}

\usepackage[table]{xcolor}
\usepackage{booktabs}
\usepackage{multirow}

\usepackage{pdflscape}

\usepackage{tikz}
\usetikzlibrary{decorations.pathmorphing, decorations.markings, calc, fadings,
decorations.pathreplacing, patterns,shapes.arrows,positioning,3d,arrows,decorations.markings,shadings,backgrounds}

\usepackage[displaymath, mathlines]{lineno}
\usepackage{framed}
\usepackage{url}
\usepackage{caption,subcaption}

\title{\sffamily Gravitational Waves from the Phase Transition of a Non-linearly Realised Electroweak Gauge Symmetry 
}
\author{Archil Kobakhidze, Adrian Manning and Jason Yue}
\affiliation[]{ARC Centre of Excellence for Particle Physics at the Terascale, School of Physics, The University of Sydney, NSW 2006, Australia}

\emailAdd{Archil.Kobakhidze@sydney.edu.au}
\emailAdd{Adrian.Manning@sydney.edu.au}
\emailAdd{Jason.Yue@sydney.edu.au}

\abstract{Within the Standard Model with non-linearly realised electroweak symmetry, the LHC Higgs boson may reside in a singlet representation of the gauge group. Several new interactions are then allowed, including anomalous Higgs self-couplings, which may drive the electroweak phase transition to be strongly first-order. In this paper we investigate the cosmological electroweak phase transition in a simplified model with an anomalous Higgs cubic self-coupling. We look at the feasibility of detecting gravitational waves produced during such a transition in the early universe by future space-based experiments.  We find that for the range of relatively large cubic couplings, $111~{\rm GeV}~ \lesssim |\kappa| \lesssim 118~{\rm GeV}$, $\sim $mHz frequency gravitational waves can be observed by eLISA, while BBO will potentially be able to detect waves in a wider frequency range, $0.1-10~$mHz. 
}

\begin{document} 
\maketitle

%%%%%%%%%%%%%%%%%%%%%%%%%%%%%%%%%%%%%%%%%%%%%
%Section
%%%%%%%%%%%%%%%%%%%%%%%%%%%%%%%%%%%%%%%%%%%%
\section{Introduction}\label{sec:intro}
After the discovery of the LHC Higgs boson, precise determination of its couplings has become imperative. Without this knowledge the nature of electroweak symmetry remains undetermined. Namely, with the current Higgs data at hand the Standard Model (SM) with a non-linearly realised $SU(2)_L\times U(1)_Y$ gauge symmetry is still a viable option%\footnote{We note that in the recent series of works \cite{Alonso:2015fsp,Alonso:2016btr,Alonso:2016oah} 
%are suggesting that the SM represents a flat manifold for the singlet scalar, and that deviations from the SM may be parameterised by introducing curvature to the manifold. Furthermore, see \cite{Grinstein:2007iv,Goertz:2014qta,Alonso:2012px,Buchalla:2013rka,Brivio:2013pma,Alonso:2012pz,Gavela:2014vra,Corbett:2015lfa,
%Hierro:2015nna,Corbett:2015mqf,Yepes:2015qwa,Yepes:2015zoa,Contino:2010mh}  for the effective theory field using a Higgs  singlet. }
 \cite{Binosi:2012cz, Kobakhidze:2012wb}. In the most economic case, the Higgs boson can be considered as an electroweak singlet particle, which admits several additional interactions beyond the conventional SM \cite{Kobakhidze:2012wb}. These new interactions, besides having interesting manifestations at the LHC and future colliders, could have played an important role in the early universe by driving a strongly first-order electroweak phase transition. Electroweak baryogenesis in this framework has been studied in Ref. \cite{Kobakhidze:2015xlz}. This paper is devoted to investigating the production of gravitational waves during the cosmological electroweak phase transition and  the feasibility of their detection in upcoming experiments.      

Nonlinearly realised electroweak gauge theory becomes strongly interacting at high energies, the famous example being $WW\to WW$ scattering in the Higgsless Standard Model. It is expected that at high energies new resonances show up, which unitarise rapid, power-law  growth of scattering amplitudes with energy in perturbation theory. However, the scale where new physics is expected to emerge crucially depends on Higgs couplings and could be as high as few tens of TeV \cite{Kobakhidze:2012wb, Kobakhidze:2015xlz}. New physics at such high energies may escape the detection at LHC. Therefore, alongside with the precision measurements of Higgs couplings, complimentary information stemming from astrophysical observations of gravitational waves may provide an important hint for the nature of the electroweak symmetry %({\color{red} Need to put reference for SKA  \cite{Kikuta:2014eja} somewhere.})
and the cosmological phase transition. 

With this motivation, we consider a simplified model with only one additional anomalous cubic Higgs coupling $\kappa$, which is the most relevant coupling for the electroweak phase transition and also one of the most difficult to be measured at the LHC. Beyond this and simplicity considerations, we have no fundamental reason to stick with this minimalistic scenario. In fact, the model can be extended in various ways, without significantly affecting our results.      
Note that, production of gravitational waves from a first order phase transition in effective theories of the SM with higher dimensional operators have been previously discussed in \cite{Delaunay:2007wb,Huang:2016odd,Leitao:2015fmj}. %Huang:2015izx,Chung:2012vg,Grojean:2004xa,Huang:2015tdv,

The paper is organised as follows. In the following section (Sec.~\ref{sec:model}) we give a brief account of the non-linear SM; The next section (Sec.~\ref{sec:fin_temp}) is devoted to discussion of the electroweak phase transition. In Sec.~\ref{sec:gw},  we compute the amplitude of gravitational waves produced during the strongly first order phase transition and identify the range of $\kappa$ for which they can potentially detected by eLISA and BBO. The conclusions are presented in Sec.~\ref{sec:conclusion} and some technical details are given in the Appendices.

%%%%%%%%%%%%%%%%%%%%%%%%%%%%%%%%%%%%%%%%%%%%%%
%%Section
%%%%%%%%%%%%%%%%%%%%%%%%%%%%%%%%%%%%%%%%%%%%%%
\section{Non-linear Realisation of the Electroweak Gauge Group}\label{sec:model}
Within the conventional SM, the linearly realised $SU(2)_L\times U(1)_Y$ electroweak gauge symmetry is a hidden gauge symmetry with a stability being the group of QED, $U(1)_Q$. Specifically, the theory in the `broken phase' is invariant under $U(1)_Q$ gauge transformations.  Therefore, to make the $SU(2)_L\times U(1)_Y$ symmetry manifest, it is sufficient to gauge only the coset group $SU(2)_L\times U(1)_Y/U(1)_{Q}$, which can be parameterised in terms of a non-linear field:
\begin{equation}
	{\cal X}(x):=e^{\frac{i}{2}\pi^i(x)T^i}
\begin{pmatrix}
	0 \\ 
	1 \\ 
	\end{pmatrix}, 	
	\label{1}
\end{equation}  
where $T^i=\sigma^{i}-\delta^{i3}\mathbb{I}$ are the three broken generators with $\sigma^i$ being the Pauli matrices. The $\pi^i (x)$ fields are the three would-be Goldstone bosons spanning the $SU(2)_L\times U(1)_Y/U(1)_{Q}$ coset space. 
With non-linear realisation of $SU(2)_L\times U(1)_Y$ electroweak gauge invariance the Higgs field $h$ is no longer obliged to form the electroweak doublet irreducible representation. In the minimal scenario  the Higgs boson resides in the $SU(2)_L\times U(1)_Y$ singlet field, $\rho(x)$. The standard Higgs doublet then can be identified with the following composite field\footnote{We note that if $\rho(x)$ field is to be identified with the modulus of the electroweak doublet field, $\rho^2=H^{\dagger}H$, it should be restricted to positive ($\rho>0$) or negative ($\rho<0$) values only.}:
\begin{equation}
H(x)=\frac{\rho(x)}{\sqrt{2}}{\cal X}(x)~.
\label{2}
\end{equation}

While maintaining $SU(2)_L\times U(1)_Y$ invariance, the non-linear realisation of the electroweak gauge symmetry allows a number of new interactions beyond those present in the SM, including anomalous Higgs-vector boson couplings, flavour and CP-violating Higgs-fermion couplings and anomalous Higgs interactions. A generic model is rather complicated and also severely constrained by the electroweak precision measurements, flavour physics and Higgs data. In this paper we only consider modification of the SM Higgs potential by adding an anomalous cubic coupling: 
\begin{equation}\label{eq:phi3_potential}
V(\rho)=-\frac{\mu^2}{2}\rho^2+\frac{\kappa}{3} \rho^3+\frac{\lambda}{4}\rho^4.
\end{equation}   

We assume that the scalar potential has a global minimum for a non-zero vacuum expectation value of the Higgs field $\rho$ (cf. the  next section): 
\begin{equation}
\langle \rho\rangle = v,~~|v|\approx 246~{\rm GeV}.
\label{eq:vev}
\end{equation}
The absolute value of the vacuum expectation value in (\ref{eq:vev}) is fixed to the standard value since the Higgs interactions with the electroweak gauge bosons are assumed to be the same as in SM, i.e.,
\begin{equation}
\frac{\rho^2}{2}(D_{\mu}{\cal X})^{\dagger}D^{\mu}{\cal X},
\label{5}
\end{equation}
where $D_{\mu}$ is an $SU(2)_L\times U(1)_Y$ covariant derivative. The shifted field
\begin{equation}
h(x)=\rho(x) - v,
\label{6}
\end{equation}
describes the physical excitation associated with the Higgs particle with the tree-level mass squared:
\begin{equation}\label{eq:h_mass}
m_h^2=\left.\frac{\partial^ 2 V}{\partial \rho\partial \rho}\right\vert_{\rho=v}\approx \left( 125~{\rm GeV}\right)^2~.
\end{equation} 
Using equations (\ref{eq:vev}) and (\ref{eq:h_mass}), we find it convenient to rewrite the mass parameter $\mu^2$ and the quartic coupling $\lambda$ in terms of the (tree level) Higgs mass, $m_h\approx 125$ GeV, the Higgs vacuum expectation value $v$  and the cubic coupling $\kappa$ as:
 \begin{eqnarray}
\mu^2& =\frac{1}{2}\left( m_h^2+v\kappa\right), \label{12a}  \\
\lambda &=\frac{1}{2v^2}\left( m_h^2-v\kappa\right). 
\label{8}
\end{eqnarray}   
The potential must also be bounded from below, that is $\lambda >0$ and, hence, $v\kappa < m_h^2$. 

The presence of the cubic term in the tree-level Higgs potential (\ref{eq:phi3_potential}) significantly alters the Higgs vacuum configuration, even without thermal corrections. With this kind of potential one can differentiate the following three cases:
\begin{itemize}
\item[(i)] A non-tachyonic mass parameter, i.e., $\mu^2<0$ or, equivalently, $v\kappa < -m_h^2$.  One of the local minima in this case is at a trivial configuration $\langle \rho \rangle=0$. We find that the electroweak symmetry breaking minimum (\ref{eq:h_mass}) is realised as an absolute minimum of the potential if $-3m_h^2<v \kappa <-m_h^2$.
\item[(ii)] A tachyonic mass parameter, i.e., $\mu^2>0$, which implies $v\kappa > -m_h^2$.  In this case the trivial configuration is a local maximum and the minimum (\ref{eq:h_mass}) is realised providing $-m_h^2<v\kappa <0$. 
\item[(iii)] For $\mu^2=0$ ($v\kappa =-m_h^2$), $v=-\frac{\kappa}{\lambda}$. In this case there are two trivial solutions for the extremum equation, which represent an inflection point of the potential.  
\end{itemize}
Notice that there exists a symmetry of the above vacuum solutions under $\kappa \to - \kappa$ and $v\to -v$. 

Although the above analysis was done at tree level, we have verified that the one-loop quantum corrections to the tree-level potential does not change the above picture significantly. For our in-depth analysis of this potential we use the one-loop thermally corrected potential which is described in the following section.

%%%%%%%%%%%%%%%%%%%%%%%%%%%%%%%%%%%%%%%%%%%%%%
%%Section
%%%%%%%%%%%%%%%%%%%%%%%%%%%%%%%%%%%%%%%%%%%%%%

\section{Bubble Dynamics in Electroweak Phase Transition}\label{sec:fin_temp}

\subsection{Finite Temperature Potential}

In order to calculate the electroweak phase transition dynamics we consider the one-loop finite temperature potential \cite{Liu:1992tn,Dine:1992wr,Arnold:1992rz,Carrington:1991hz}. The potential, as a function of temperature, $T$, can be split into the following parts:
\begin{equation}\label{eq:finitepotential}
  V(\rho,T)=  V^{(0)}(\rho)+ V^{(1)}_{CW}(\rho) + V^{(1)}(\rho,T>0) + V_{Daisy}(\rho,T>0),
\end{equation}
where $V^{(0)}$ is the classical potential and is given in (\ref{eq:phi3_potential}), $V_{CW}^{(1)}$ is the Coleman-Weinberg contribution for $T=0$ and is given by%\footnote{This corresponds to choosing the renormalisation condition the quantum correction at $\rho_*$ does not shift the tree level potential at the point i.e. $V_{CW}(\rho_*)=0$ where $m^2_h(\rho_*)=0$ or $v^2 e^{3/2}$. Otherwise if one impose $V_{CW}(v)=0$ then one recovers 
%\begin{equation}
%    V_{CW}(\rho) = \sum_{i=W,Z,t,h} n_i \frac{m_i^4(\rho)}{64 \pi^2}\left[\left(\ln\left(  \frac{ m^2_i(\rho)}  {m_i^2(v)}\right)- \frac{3}{2}   
%    \right)+2m_i^2(\rho)m_i^2(v)-\frac{1}{2}m_i^2(v)\right],
%\end{equation} from \cite{Leitao:2015fmj}.}
:
      \begin{equation}
        V_{CW}^{(1)} (\rho) = \sum_{i=W,Z,t,h} n_i \frac{m_i^4(\rho)}{64 \pi^2}\left(\ln\left(  \frac{ m^2_i(\rho)}  {v^2}\right)- \frac{3}{2} \right) ~.
        \end{equation}
        $V^{(1)}(\rho,T>0)$ is the finite temperature contribution and is defined via the thermal function $J$:
\begin{equation}\label{eq:thermal}
  \begin{aligned}
    V^{(1)}(\rho,T)&= \frac{T^4}{2\pi^2} \sum_{i=W,Z,t,h} n_i J\left[\frac{m_i^2(\rho)}{T^2}\right],\\
    J[m^2_i\beta^2] & :=\int^\infty_0 dx \ x^2 \ln \left[1-(-1)^{2s_i+1}e^{-\sqrt{x^2+\beta^2m^2_i}}\right],
  \end{aligned}
\end{equation}
here $s_i$ corresponds to the spin and $n_i$ to the number of degrees of freedom of the particle species $i$. $V_{Daisy}(\rho,T>0)$ are Daisy-corrected terms (see \ref{app:daisy}).

\begin{figure}[h!]
    \centering
    \begin{subfigure}[h]{0.49\textwidth}
    \includegraphics[width=\textwidth]{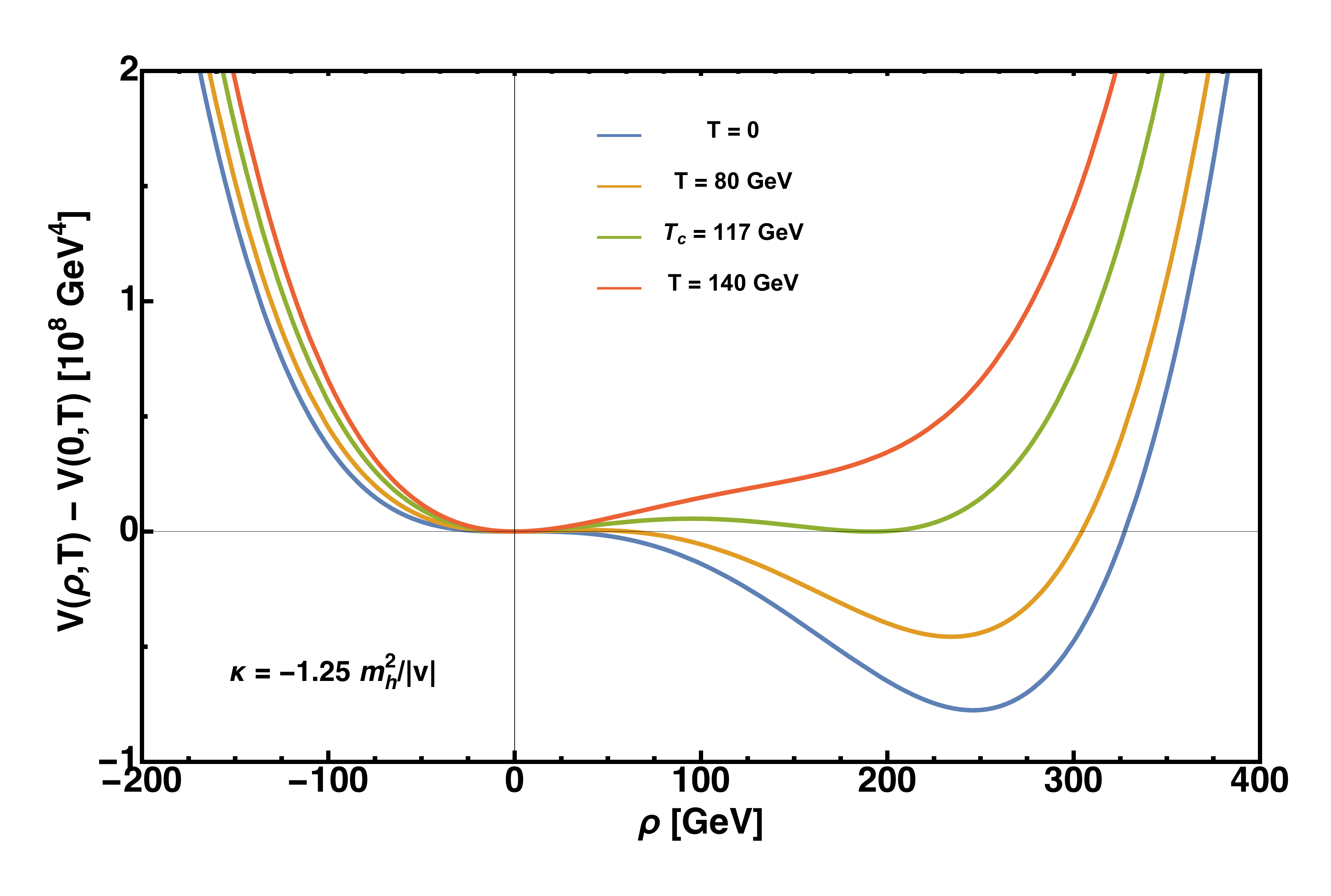}%
    \caption{}\label{fig:p125}
    \end{subfigure}
    \begin{subfigure}[h]{0.49\textwidth}
    \includegraphics[width=\textwidth]{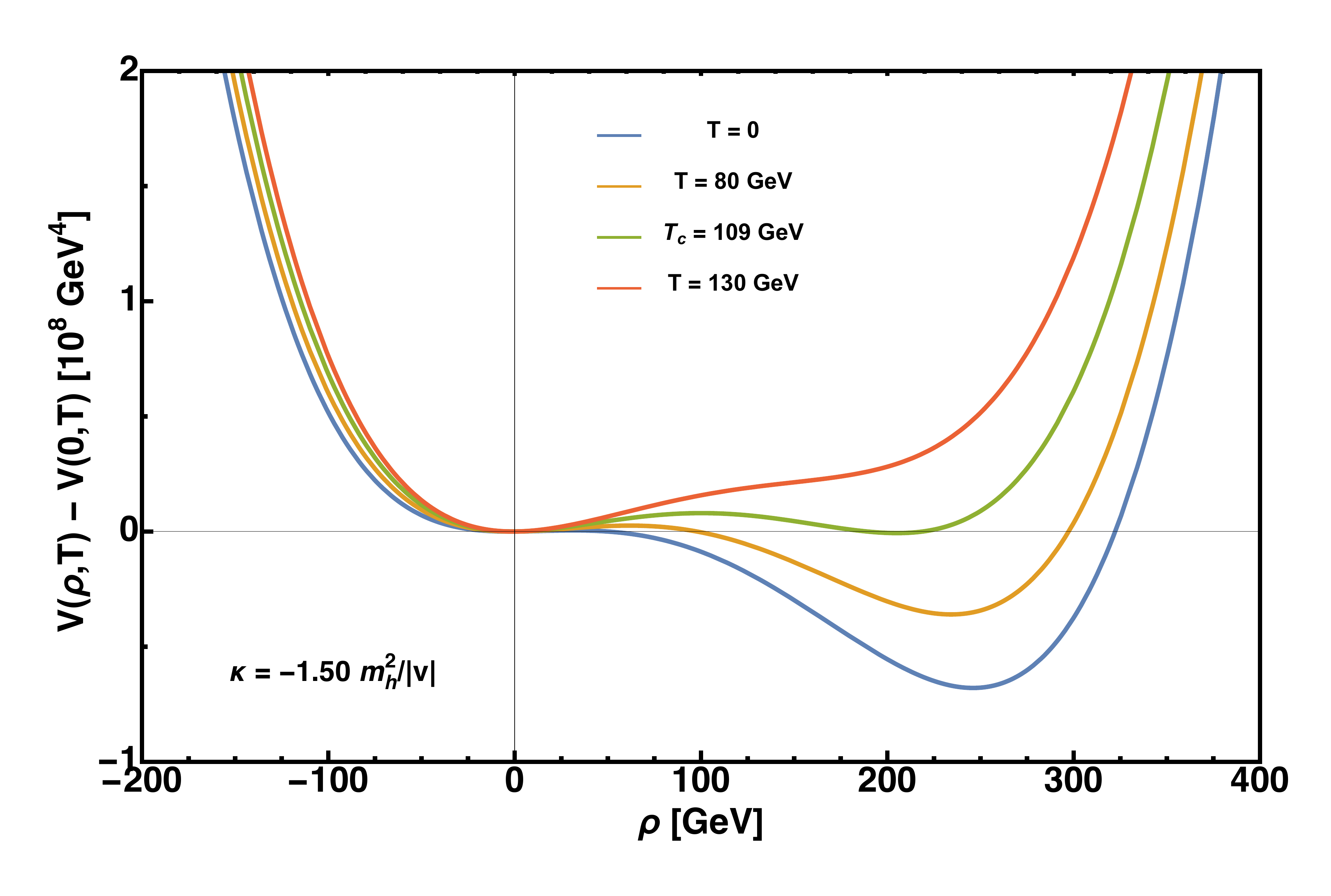}
    \caption{}\label{fig:p15}
  \end{subfigure} \\[0.1cm]

    \begin{subfigure}[h]{0.49\textwidth}
    \includegraphics[width=\textwidth]{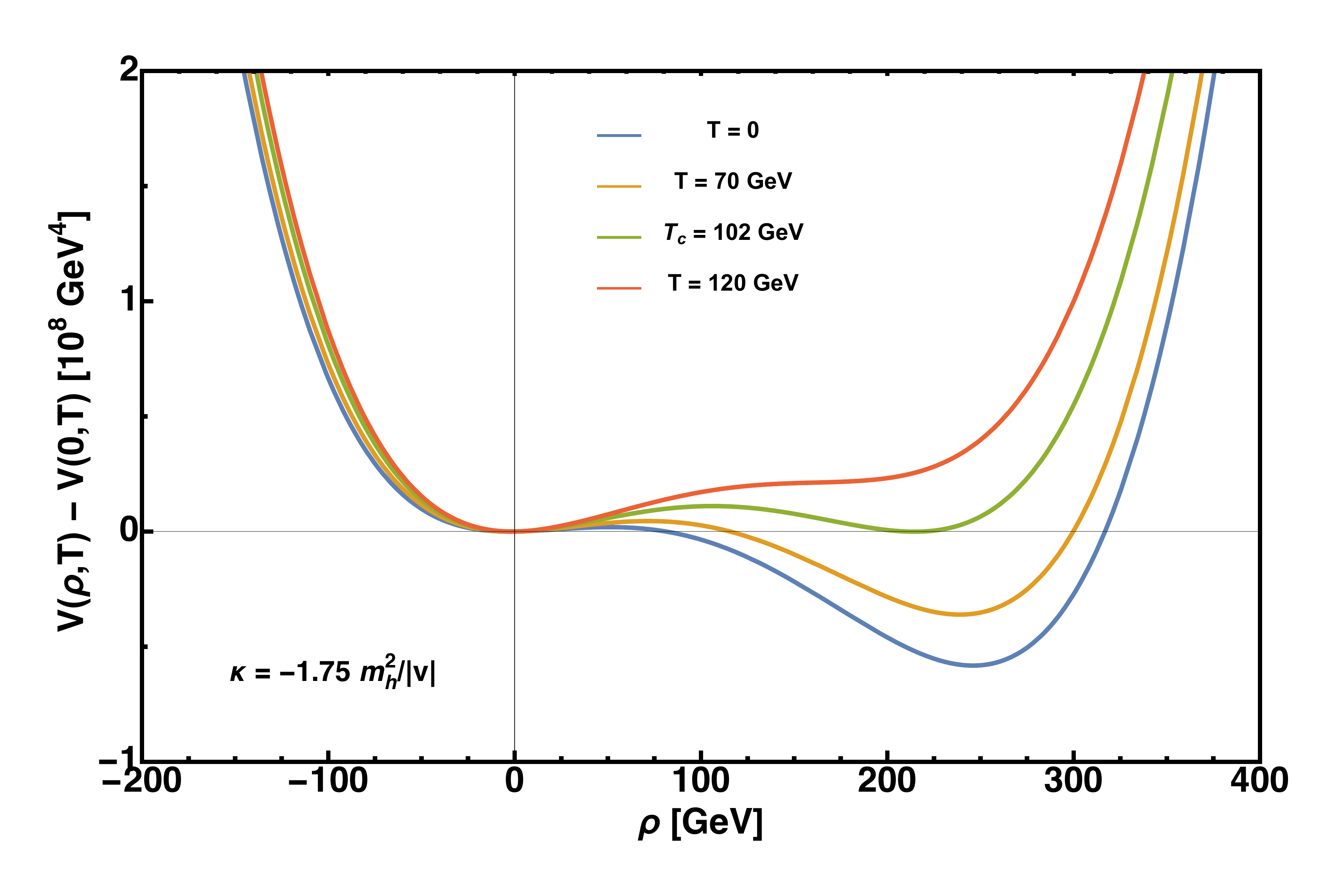}%
    \caption{}\label{fig:p175}
    \end{subfigure}
    \begin{subfigure}[h]{0.49\textwidth}
    \includegraphics[width=\textwidth]{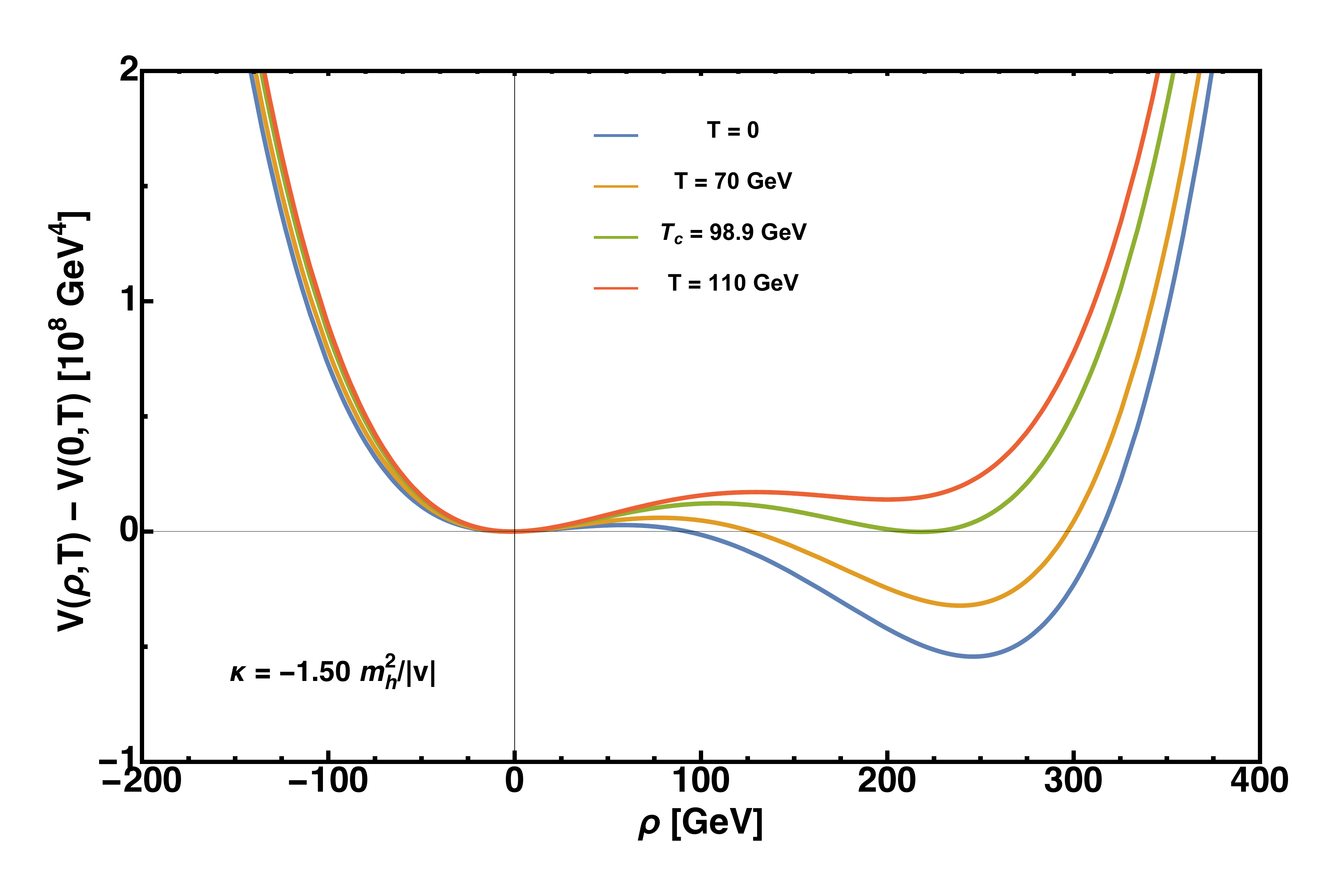}
    \caption{}\label{fig:p185}
    \end{subfigure}
    \caption{ The finite temperature effective potential for cubic couplings taking the values (\subref{fig:p125}) $\kappa = -1.25 m_h^2/|v|$,  (\subref{fig:p15})  $\kappa=-1.50 m_h^2/|v|$, (\subref{fig:p175}) $\kappa=-1.75 m_h^2/|v|$, (\subref{fig:p185}) $\kappa =-1.85 m_h^2/|v|$. The potential are shown for temperatures above and below that of $T_c$, as well as $T_c$ itself.  }
    \label{fig:results_pot}
\end{figure}

Although this contribution can be expanded for high temperatures, it has previously been shown in \cite{Kobakhidze:2015xlz} that an anomalous cubic term in the potential may lower the critical temperature down to $\sim 50$ GeV, rendering the high temperature expansion invalid for this work. We therefore numerically evaluate the full form of (\ref{eq:thermal}) in all following calculations. The total potential is shown for a range of $\kappa$ values and temperatures in Fig \ref{fig:results_pot}.

There exists IR divergences when the field-dependent Higgs mass as defined via (\ref{eq:h_mass}), % $m_h^2(\rho)=-\mu^2+2\kappa\rho+3 \lambda\rho^2 $
(at one-loop) becomes negative. The inclusion of the Daisy-corrections (cf. \ref{app:daisy}) shrink the region of instability, however there are regions of $\rho$ at given temperatures where these divergences remain. For this reason, we impose a temperature dependent cut-off as a lower bound in the integral (\ref{eq:thermal}) which corresponds to the minimum momentum required such that the integral is real for all $\rho$ values. Note also that the $\rho$ mass completely specifies the stability of the vacuum in the SM whereas this is not the case for a non-linearly realised case.

\subsection{Nucleation Temperatures}

A salient feature of the effective finite temperature potential (\ref{eq:finitepotential}) is that it contains a single minimum at high temperatures at non-zero values of the Higgs field. Hence, the SM field remains massive in the high-temperature limit.   We define this temperature-dependent vacuum state as the false vacuum, $v_T^{(+)}$. As the universe cools, this potential develops a secondary minimum, which decreases with temperature. Below the critical temperature, $T_c$, defined as the temperature where the two minima become degenerate, tunnelling from the false vacuum to the secondary minimum (true vacuum $v_T^{(-)}$) becomes energetically favourable. The cubic addition to this potential makes this phase transition strongly first order and therefore result in electro-weak bubble formation in the early universe. The tunnelling probability of this phase transition per four volume, $\Gamma$, is given by:
\begin{equation}\label{eq:tun_rate}
  \Gamma(T) = \left(\frac{S_3(T)}{2\pi T }  \right)^{3/2}   T^4  e^{-S_3(T)/T}, 
\end{equation}
%where a formal  equivalence between finite temperature theory and the vacuum theory may be obtained by compactifying the imaginary time direction $\tau$ with periodicity $2n\pi T$, $n\in \mathbb{Z}$.
where we have introduced the three-dimensional Euclidean action $S_3(T)$, which is related to that in four dimensions via  $S_4[\rho]=S_3[\rho]/T$. To calculate the field profile of a nucleating bubble we must assume an $O(3)$ symmetry
%\footnote{In fact it is proven in \cite{Coleman:1977th} that any solution $\rho\neq\rho(r)$ always leads to a higher bounce action}
in the $\rho$ field. With this assumption, the Euclidean action has the form \cite{Linde:1981zj,Linde:1977mm,Linde:1980tt}:
\begin{equation}\label{eq:3d_action}
  S_3[\rho]=4\pi \int^\infty_0 dr\ r^2 \left[ \frac{1}{2}\left( \frac{d\rho}{dr} \right)^2+ V(\rho,T)-V\Big(v_{T}^{(+)},T\Big) \right], 
\end{equation}
where the potential difference may be defined as the free energy density $\mathcal{F}(\rho,T):=V(\rho,T)-V\Big(v_{T}^{(+)},T\Big)$. 
The equation of motion for this action is given by:
\begin{equation}  \label{eq:eom_bounce}
  \frac{d^2\rho}{dr^2}+\frac{2}{r}\frac{d\rho }{dr}-\frac{\partial \mathcal{F}}{\partial \rho}(\rho,T)=0~.
\end{equation}

%Physically, the formation of bubbles have competing effects --- the surface tension (given by the derivative terms in Eq. (\ref{eq:3d_action})) acts to collapse the bubble, whereas the free energy density sustains the bubble expansion. There exists a critical solution to the equation of motion which corresponds collapse nor expand, known and its  equation of motion is given by:

\begin{figure}[h]
    \centering
    \includegraphics[width=0.85\textwidth]{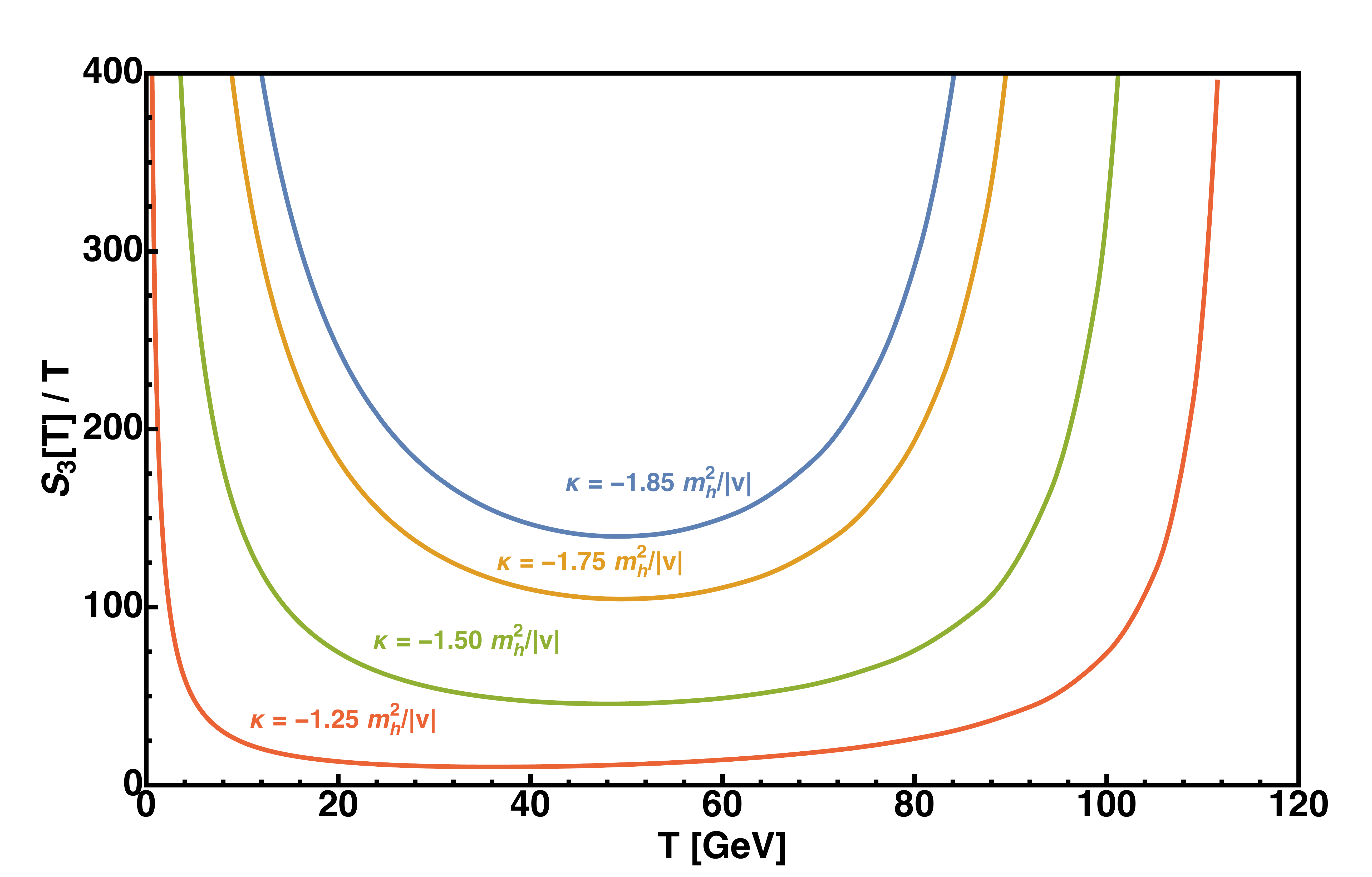}
    \caption{The numerical results for the Euclidean action vs temperature for applicable values of cubic couplings, $\kappa$.   \label{fig:action}} 
\end{figure}

The field configuration which corresponds to the field tunneling from the false vacuum to the true vacuum is known as the bounce configuration. The appropriate boundary conditions for this are:
\begin{equation}
  \frac{d\rho}{dr}(0,T)=0, \qquad \lim_{\rho\rightarrow \infty}  \rho(r,T)= v_{T}^{(+)},
  \label{eq:eom_boundary}
\end{equation}
where the latter is also necessary for the finiteness of $S_3(T)$. The solutions to (\ref{eq:eom_bounce}) and (\ref{eq:eom_boundary}) are found numerically using the shooting method \cite{Press:2007,Apreda:2001us}. The field profile obtained can be integrated to give a value for the Euclidean action as prescribed in (\ref{eq:3d_action}). These numerical values have been plotted for various cubic coupling $\kappa$ are shown in Fig.~\ref{fig:action}. 

%Moving to a cosmological setting, the scale factor in radiation domination is given by $a(t) \sim t^{1/2}$ ({\color{red} specify whether this is comoving or conformal time})and so 
%\begin{equation}
%    H=\frac{\dot{a}(t)}{a(t)}=\frac{1}{2t}=\left(\frac{ 4 \pi^3 g_*}{45}\right)^{1/2} \frac{T^2}{ M_{P}}
%  \end{equation}
%  giving the temperature-time relation:
%  \begin{equation}
%      t=0.30 \frac{M_{P}}{g_*^{1/2}T^2}\sim \left( \frac{1 \ \text{MeV}}{T} \right)^2\ \text{s}
%    \end{equation}

Typically the temperature at which an $O(3)$ symmetric bubble can form is lower than the critical temperature. This means that the onset of the phase transition can be significantly delayed. Ultimately, we seek the temperature at which we expect at least one bubble to form in our Hubble horizon. We define this temperature as the nucleation temperature, $T_n$. More explicitly, this temperature corresponds to the time required for the probability of nucleating a bubble in the Hubble horizon to be of order 1 and can be calculated using\cite{Hogan:1984hx,Anderson:1991zb,Enqvist:1991xw,Turner:1992tz} (see also \cite{Caprini:2011uz,No:2009}):
%This supercooling phenomenon can cause the onset of the phase transition to be significantly delayed. 
\begin{equation}\label{eq:nucl_temp}
    P% \sim\int_0^{t_*} \Gamma \mathcal{V}_H dt 
    \sim\int_{T_n}^{T_c}   \left( \sqrt{\frac{45}{4\pi^3g_*}}\frac{M_{P}}{T} \right)^4 e^{-S_3/T}\frac{dT}{T} \sim \mathcal{O}(1).
\end{equation}
where we have implicitly used the time-temperature relation $dT/dt=-TH$
by assuming an adiabatic expansion of the universe. 

It is easy to see from Fig.~\ref{fig:action} that the exponent in (\ref{eq:nucl_temp}) namely $S_3/T$, decreases with $|\kappa|$. For lower $S_3/T$ values, the larger the integrand in (\ref{eq:nucl_temp}) and the closer the nucleation temperature $T_n$ becomes to the critical temperature, $T_c$. We find that for $|\kappa|$ values less than $\sim1.25 m_h^2/|v|$, the nucleation and critical temperatures become degenerate. This is traced back to the fact that the barrier between the two minima becomes smaller and the strength of the first order phase transition decreases. We therefore expect the contribution to gravitational waves from these parameters to be small and focus on higher $|\kappa|$ values. Conversely, for larger $|\kappa|$ values, around $\gtrsim 2 m_h^2/|v|$ the action is large, implying that the tunneling rate is small. For these values we find that a bubble will never nucleate in a Hubble volume and will therefore leave the universe trapped in the high temperature phase. We therefore focus on $|\kappa|$ values in the interval $[1.25,1.85] \times m_h^2/|v|$.
% As a general trend, the nucleation temperatures are observed to decrease with increasing $|\kappa|$ values.

To illustrate these trends we have tabulated the various important temperatures and minima related to the phase transition for a range of cubic couplings (see Tab.~\ref{tab:T_c}).
%The temperatures (i) $T_c$ at which the potential has two degenerate minima; (ii)  $T_n$ at which nucleation effectively begins (cf. Eq.~\ref{eq:nucl_temp}) and (iii) $T_*$ at which the formation of spherical critical bubbles begins, are given in Tab.~\ref{tab:T_c} for a range of cubic couplings.  
%%%%%%%%%%%%
%%Table
%%%%%%%%%%%%
%%%%%%%%%%%%%%%%%%%%%%%%%%%%%%%%%%%%%%%%%%%%%%
\begin{table}[h!]
\begin{center}\renewcommand{\arraystretch}{1.2}
\begin{tabular}[c]{c| c| c| c| c| c}
  \hline
  $\kappa$ [$m_h^2/|v|$] & $T_n$ [GeV] & $T_*$ [GeV] & $T_c$ [GeV] & $v_{T_c}^{(-)}$ [GeV] & $v_{T_c}^{(+)}$ [GeV] \\ \hline\hline
\rowcolor{gray!20}
$-1.00$   & 115. & 115.  & 125.  & $0.625$  & 175.    \\ \hline  
$-1.25$   & 106. & 111.  & 117.  & $-0.430$ & 191.    \\ \hline  
\rowcolor{gray!20}
$-1.50$   &  91.9 & 102.  & 109.  & $-1.09$  & 204.    \\ \hline  
$-1.75$   & 72.0  & 91.3    & 102.  & $-1.52$  & 214.    \\ \hline  
\rowcolor{gray!20}
$-1.85$   & 56.1  & 86.8    & 98.9  & $-1.68$  & 218.    \\ \hline  
$-2.00$   &  -     & 79.0    & 94.3  & $-1.88$  & 223.    \\ \hline       
\rowcolor{gray!20}
$-2.50$   &  -     & 44.7   & 77.1  & $-2.42$  & 235.    \\ \hline    
\end{tabular}                                                        
\end{center}                                                         
\caption{Critical temperatures and the field configurations of the degenerate potential. $T_n$ is the temperature at which the probability of forming a spherical bubble in a Hubble horizon is of order 1. $T_*$ is the temperature at which spherical bubbles can begin to nucleate. $T_c$ is the critical temperature, where the two vacua are degenerate. $v_T^{(\pm)}$ are the false and true vacua respectively.}\label{tab:T_c}
\end{table}
%%%%%%%%%%%%%%%%%%%%%%%%%%%%%%%%%%%%%%%%%%%%%%                       

%%%%%%%%%%%%%%%%%%%%%%%%%%%%%%%%%%%%%%%%%%%%%%                       
%%Section
%%%%%%%%%%%%%%%%%%%%%%%%%%%%%%%%%%%%%%%%%%%%%%
\section{Predictions for Gravitational Waves}\label{sec:gw}

%For first order electroweak phase transitions in the early universe, it is expected that interactions between nucleated bubbles will generate gravitational waves (for reviews of gravitational waves and their detection, cf. \cite{Binetruy:2012ze,Maggiore:1999vm,Buonanno:2007yg}). 

During a first-order phase transition in the early universe, gravitational waves are known to be generated from three major production mechanisms. Collisions of bubble walls \cite{Huber:2008hg}, sound waves in the plasma which occur after bubble collisions \cite{Hindmarsh:2015qta} and magnetohydrodynamical turbulence which can also form in the plasma post bubble collision \cite{Binetruy:2012ze,Caprini:2009yp}. We will systematically address each of these contributions in this section.

In the following discussion will be expressing the energy density of gravitational wave radiation in terms of the quantity,
\begin{equation}
  h^2 \Omega_{GW}(f) = \frac{h^2}{\rho_c} \frac{d \rho_{GW}}{d (\ln f)} ~,
  \label{eq:gw_ed}
\end{equation}
where $\rho_{GW}$ is the gravitational wave energy density, $f$ is the frequency and $\rho_c= 3H_0^2/(8\pi G) $ is the critical energy density today. Here $h_0$ is the Hubble rates measured in units of $100$ s$^{-1}$       Mpc$^{-1}$.
% Pretty sure people know what the Hubble rate is, we dont need to define it and find references. 
%The Hubble rate  usually re-expresssed in terms of $H_0=100 h_0$ km(sec-Mpc)$^{-1}$ where $h_0 \in [0.5,0.85]$ (PDG 1994 {\color{red} Need more update reference}) parameterises the experimental uncertainties  \cite{Maggiore:1999vm}.

The total energy density of produced gravitational waves can be written as a sum of the three main production mechanisms (as they approximately linearly combine\cite{Caprini:2015zlo}):
\begin{equation}\label{eq:gw_contri}
  h^2 \Omega_{\text{GW}} \simeq h^2 \Omega_{col} + h^2 \Omega_{sw} +h^2 \Omega_{\text{MHD}}~. 
\end{equation}
These contributions are, in order of their appearance in (\ref{eq:gw_contri}), bubble collisions, sound waves and magnetohydrodynamic turbulences\footnote{For discussion of scalar field potential of form (\ref{eq:phi3_potential}) interacting with relativistic fluid, see \cite{Giblin:2014qia,Giblin:2013kea}.}.

The gravitational wave spectrum corresponding to these contributions are conveniently parameterised in terms of two parameters which therefore also characterises the phase transition.  

The first parameter, $\alpha$, is the ratio of the latent heat released by the phase transition normalised against the radiation density:
    \begin{equation}
        \begin{aligned}
              \alpha&:=\frac{\epsilon_*}{\rho_{rad}}=\frac{1}{\frac{\pi^2}{30}g_*T_*^4}\left( -\Delta V +T_*\Delta s \right)\\
              \Delta V& = V\left(v_{T_*}^{(-)},T_*\right)-V\left(v_{T_*}^{(+)},T_*\right),\\
              \Delta s& =  \frac{\partial V}{\partial T} \left(v^{(-)}_{T_*},T_*\right)
                        -  \frac{\partial V}{\partial T}\left (v^{(+)}_{T_*},T_*\right). 
%                  &=\frac{1}{\frac{\pi^2}{30}g_*T_*^4}\left[\left( -V(T_*,v^{(-)}_{T_*})  + T_* \frac{d}{dT} V(T_*,v^{(-)}_{T_*})\right) - \left( -V(T_*,v^{(+)}_{T_*})  +    T_* \frac{d}{dT} V(T_*,v^{(+)}_{T_*})\right) \right]
                    \end{aligned}
                  \end{equation}
For non-run-away bubbles, the energy released grows as the size of $\sim R^3$ but the kinetic energy of the bubble wall scales as its surface area $\sim R^2$. In this scenario, a large fraction of the energy goes into the reheating and fluid motion of the plasma, which corresponds to the last two sources in (\ref{eq:gw_contri}). 

However the energy that can be deposited into the fluid saturates at:
\begin{equation}\label{eq:alpha_runaway}
  \begin{aligned}
    \alpha_\infty&: =\frac{30}{\pi^2} \left({\displaystyle\sum_{i\in\{t,W,Z\}}c'_i n_i }\right)^{-1} \left({\displaystyle\sum_{i\in\{t,W,Z\}} c_i n_i \Delta y_i^2}\right)
    \left( \frac{v_{T_*}^{(-)}}{T_*} \right)^2 \\
     & \approx 4.9\times 10^{-3} \left( \frac{v_{T_*}^{(-)}}{T_*} \right)^2,
  \end{aligned}
\end{equation}
where $c_i=1\ (1/2)$, $c'_i=1(7/8)$ for bosons (fermions), and $y_i$ are the coupling strengths to the Higgs boson. We note that although the Higgs contributions with an anomalous term will no longer be proportional to $\left|v_{T_n}^{(-)}\right|^2$ but its inclusion to (\ref{eq:alpha_runaway}) will modify the bound to at most 5\%. For $\alpha >\alpha_\infty$, the bubble wall will accelerate indefinitely (until it reaches $v=1$). 
The fraction of surplus energy which is converted into wall acceleration is given by:
\begin{equation}
  \kappa_\rho=  1-\frac{\alpha_\infty}{\alpha}. 
\end{equation}
The fraction that goes into bulk motion is then:
\begin{equation}\label{eq:gw_eff}
  \kappa_v=\frac{\alpha_\infty}{\alpha} \underbrace{\left(\frac{\alpha_\infty}{0.73+0.083\sqrt{\alpha_\infty}+\alpha_\infty} \right)}_{:=\kappa_\infty}.
  \end{equation}

For the values of $|\kappa|\in [1.25, 1.85] \times m_h^2/|v| = [79 ,118]$ GeV considered here, the expansion of the bubble wall always dominates over the friction due to the surrounding plasma, ensuring $\alpha > \alpha_\infty$. We therefore expect for this range of $\kappa$ values, the bubble walls will accelerate without bound resulting in what is known as the run-away configurations \cite{Leitao:2015fmj,Megevand:2009gh,Bodeker:2009qy}.
%We are therefore concerned with thermal or vacuum run-aways \cite{Hindmarsh:2015qta,Caprini:2015zlo} which differs in the energy budget of the transition \cite{Espinosa:2010hh}. 

The second parameter characterising the gravitational wave spectrum is $\beta$. It is the inverse time of the phase transition and can be defined through: 
\begin{equation}
   \Gamma(t)=\Gamma(t_*) e^{-\beta (t-t_*)+\ \ldots} \qquad \Longleftrightarrow 
   \qquad  \frac{\beta}{H_*}=T_* \left.\frac{d}{dT}\right|_{T=T_*}\left(\frac{S_3(T)}{T}  \right),
\end{equation}
where the reference time and temperature are usually chosen to that at nucleation, so that $T_* = T_n$, $t_*=t_n$ and $H_*=H(T_n)$.  

%Typical estimates give the duration of the phase transition (from nucleation to bubble collision) as $\Delta t \sim \beta^{-1}$ so that  bubble sizes are of order $v_w \beta^{-1}$ \cite{Kosowsky:1992vn}, where $v_w$ is the bubble wall velocity. However, for the strong phase transitions  that was found in this work, the bounce action quickly varies with temperatures and may have  significant departures from $\beta/H\sim S_3/T\sim \mathcal{O}(10^2)$ (cf. Fig.~\ref{fig:action}).

In Tab.~\ref{tab:alpha_beta}, we show the typical values of $\alpha$ and $\beta$ for the range of cubic coupling we consider. Similarly to the values found in \cite{Kehayias:2009tn,Leitao:2015fmj} we see the trends that as the cubic coupling increases, the strength of the first-order phase transition increases, the latent heat ($\propto \alpha$) increases and the duration of the transition ($\propto \beta^{-1}$) increases.

%These values lie in the range found in   and it for larger  $\kappa$ values leading to stronger phase transitions, the false vacuum energy (as parameterised by) $\alpha$ whereas the timescale (as parameterised by $\beta$ decrease.
%  for zero temp case see --- Adams:1993zs

%%%%%%%%%%%%
%%Table
%%%%%%%%%%%%
%%%%%%%%%%%%%%%%%%%%%%%%%%%%%%%%%%%%%%%%%%%%%%
\begin{table}
\begin{center}\renewcommand{\arraystretch}{1.3}
\begin{tabular}[c]{c| c| c| c| c }
\hline
$\kappa$ $[m_h^2/|v|]$& $T_n$ GeV & $\alpha$& $\beta/H_n$ & $v_{T_n}^{(-)}/T_{n}$  \\ \hline
%1.0 & 114.81 & 0.028 & Too Sensitive &   1.4 & 0.70 \\ \hline 
\rowcolor{gray!20}
$-1.25$ & 106. & 0.037 & 1770 &  1.64 \\ \hline 
$-1.50$ & 91.9 & 0.057 & 989. &  1.87 \\ \hline 
\rowcolor{gray!20}
$-1.75$ & 72.0 & 0.11 & 308.  &  2.11 \\ \hline 
$-1.85$ & 56.1 & 0.24 & 69.5  &  4.33 \\ \hline 
\rowcolor{gray!20}
$-2.00$ & - & - &  - &  - \\ \hline
\end{tabular}
\caption{Summary of gravitational wave parameters for a range of cubic coupling, $\kappa$'s. $\alpha$ is the latent heat divided by radiation density, $\beta$ is the inverse tunneling rate and $v_{T_n}^{(-1)}/T_n$ is a measure of the strength of the first order phase transition.\label{tab:alpha_beta}}
\end{center}
\end{table}
%%%%%%%%%%%%%%%%%%%%%%%%%%%%%%%%%%%%%%%%%%%%%%

%With the gravitational waves effectively decoupled from the universe, the energy density 

%scales as $a(t)^{-4}$ and the frequency as $a(t)^{-1}$. The adiabicity condition implies that the entropy density $s\sim a(t)^3 g(T) T^3$ remains constant and so: 
%\begin{equation}
%    \frac{a_*}{a_0}= \frac{a(t_*)}{a(t_0)} = 8.0 \times 10^{14} \left( \frac{100}{g_*} \right)^{1/3} \left(\frac{1\ \text{GeV}}{T_*}  \right).
%  \end{equation}

The gravitational waves effectively decouple from the universe, meaning the energy density and characteristic frequencies will need to be red-shifted to compare to the values measured today, specifically (cf. e.g. \cite{Kamionkowski:1993fg}):
\begin{equation}
  \begin{aligned}
    f_0 &= f_* \left( \frac{a_0}{a_*} \right)
    &&=1.65\times 10^{-7} \ \text{Hz} \left( \frac{f_*}{H_*}\right)  \left( \frac{T_*}{1\ \text{GeV}} \right)
           \left( \frac{g_*}{100} \right)^{1/6},\\
    \Omega_{GW} &= \Omega_{GW*} \left( \frac{a_0}{a_*} \right)^4 \left( \frac{H_*}{H_0} \right)^2
    &&= 1.67\times 10^{-5} h_0^{-2}\left( \frac{100}{g_*} \right)^{1/3} \Omega_{GW*}.
  \end{aligned}
\end{equation}
This red-shifting moves the peak frequency of gravitational waves from early phase transitions to the mHz region, requiring space-based detectors to measure.
The analytical fits to the gravitational wave energy spectra from the various sources will be summarised in the subsequent section. 

\subsection{Field Profiles}
For bubble collisions, due to the Higgs scalar field profile, the gravitational wave spectrum can be approximated by the envelope approximation \cite{Huber:2008hg} giving:
\begin{equation}\label{eq:coll}
  h^2 \Omega_{col }(f) =  1.67\times 10^{-5} \left(\frac{\beta}{H_n} \right)^{-2} \left( \frac{\kappa_v \alpha }{1+\alpha} \right)^2 
  \left( \frac{100}{g_*} \right)^{1/3}\left( \frac{0.11v_w^3}{0.42+v_w^2} \right)
S_\rho(f),
\end{equation}
where 
\begin{equation}
  \begin{aligned}
    S_{\rho}(f)& =\frac{3.8 (f/f_\rho)^{2.8}}{1+2.8(f/f_\rho)^{3.8}},\\ 
    f_{\rho} &= 16.5 \times 10^{-3} \left( \frac{0.62}{1.8-0.1v_w +v_w^2} \right) \left( \frac{\beta}{H_n} \right)\left( \frac{T_n }{100 \text{ GeV}} \right)\left( \frac{g_*}{100} \right)^{1/6}\ \text{mHz}.
  \end{aligned}
\end{equation}
%{\color{red} Here, the characteristic frequency is set by the duration of the transition $\sim \beta$ to give the first factor in $f_\rho$. \ldots }

Bulk motion in the fluid after bubble collisions is known to provide acoustic production of gravitational waves. This sound wave contribution to the gravitational wave spectrum is estimated by \cite{Hindmarsh:2015qta,Hindmarsh:2013xza}: 
\begin{equation}\label{eq:sw}
  h^2 \Omega_{sw}(f) =  2.65\times 10^{-6} \left(\frac{\beta}{H_n} \right)^{-1} \left( \frac{\kappa_v \alpha }{1+\alpha} \right)^2 
  \left( \frac{100}{g_*} \right)^{1/3}v_w S_{sw}(f),
\end{equation}
where the spectral shape and characteristic frequency are given by:
\begin{equation}
  \begin{aligned}
    S_{sw}(f)& = \left( \frac{f}{f_{sw}} \right)^3\left( \frac{7}{4+3(f/f_{sw})^2} \right)^{7/2},\\
    f_{sw} &= 1.9 \times 10^{-2} v_w^{-1} \left( \frac{\beta}{H_n} \right)\left( \frac{T_n }{100 \text{ GeV}} \right)\left( \frac{g_*}{100} \right)^{1/6}\ \text{mHz}.
  \end{aligned}
\end{equation}
These results are fitted from numerical simulation for generic values of $v_w \lesssim 0.9$. As the plasma in the early universe is fully ionized, magnetohydrodynamical turbulence can give rise to gravitational waves after bubble collisions. The contribution to the gravitational wave spectrum can be approximated by \cite{Caprini:2009yp}:
\begin{equation}\label{eq:mhd}
  h^2 \Omega_{\textsc{MHD}}(f) =  3.35\times 10^{-4} \left(\frac{\beta}{H_n} \right)^{-1} \left( \frac{\epsilon \kappa_v \alpha }{1+\alpha} \right)^{3/2} 
  \left( \frac{100}{g_*} \right)^{1/3}v_w S_{\textsc{MHD}}(f),
\end{equation}
where $\epsilon\sim 5-10\% $ and we take the lower value for more conservative estimate. The spectral shape and characteristic frequency for this contribution are:
\begin{equation}
  \begin{aligned}
    S_{\textsc{MHD}}(f)& = \left( \frac{f}{f_{\textsc{MHD}}} \right)^3\left( \frac{1}{\left[ 1+(f/f_\textsc{MHD}) \right]^{11/3} (1+8\pi f /h_n)} \right),\\
    f_{\textsc{MHD}} &= 2.7\times 10^{-2}  v_w^{-1} \left( \frac{\beta}{H_n} \right)\left( \frac{T_n }{100 \text{ GeV}} \right)\left( \frac{g_*}{100} \right)^{1/6}\ \text{mHz}.
  \end{aligned}
\end{equation}
Differing from the previous two contributions, there is an explicit dependence of the spectral shape on the Hubble rate via $h_n$. This together with the extra $\beta/H$ factor for  turbulence (cf. Eq.~\ref{eq:mhd})  and  sounds wave (cf. Eq.~\ref{eq:sw})  reflects that fluid motion is typically a last longer source of gravitational waves than bubble collisions (cf. Eq.~\ref{eq:coll}). 

%As discussed in the previous part, the timescale of the phase transition is usually taken as $\Delta t\sim \beta^{-1}$. Even when the better approximation $\Delta t \approx 3 \beta^{-1}\ln \left( \beta/H \right) $ was used to track the phase transition up to percolation, and to account for substantially deviations from $\beta/H\sim S_3/T$,  the turbulence contribution was found to be less significant than other sources \cite{Leitao:2012tx,Leitao:2015fmj}. 

Now, we move on to discuss the detection of these sources as generated  a first order phase transition of a our non-linearly realised EW gauge group. 
\subsection{Detection at Next Generation Space-Based Interferometers}

\begin{figure}[h]
    \centering
    \begin{subfigure}[h]{0.49\textwidth}
    \includegraphics[width=\textwidth]{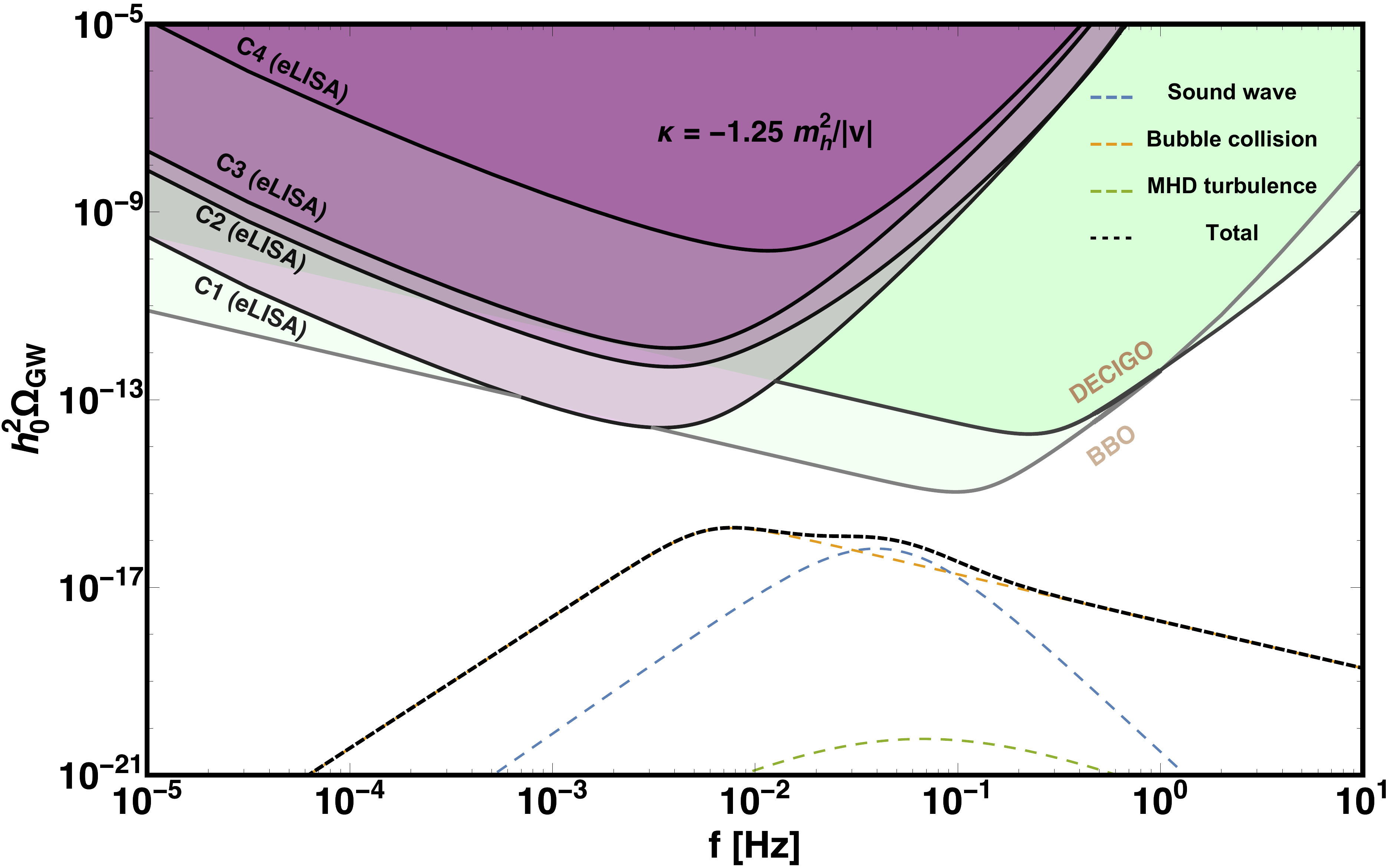}%
    \caption{}\label{fig:k125}
    \end{subfigure}
    \begin{subfigure}[h]{0.49\textwidth}
    \includegraphics[width=\textwidth]{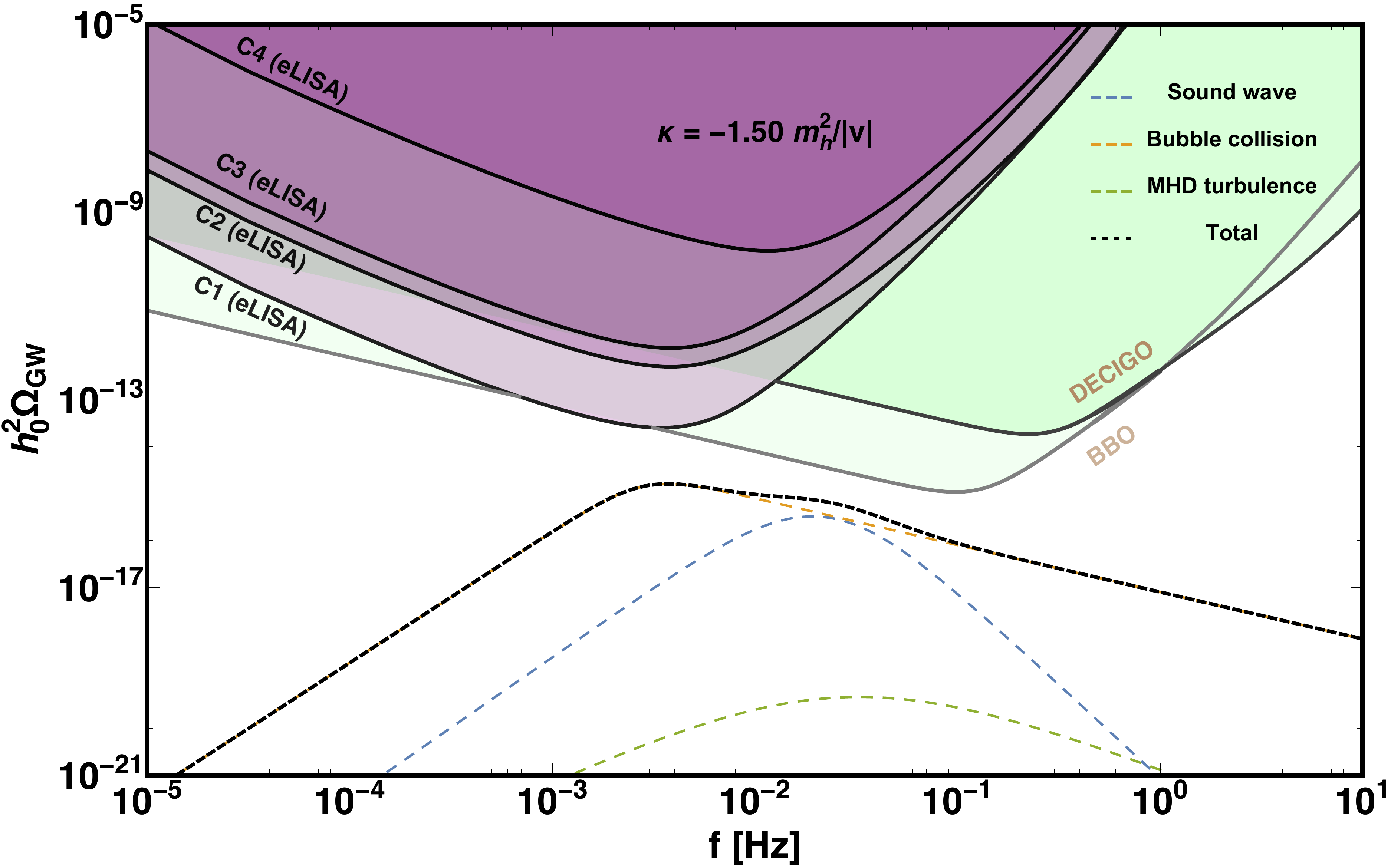}
    \caption{}\label{fig:k15}
  \end{subfigure} \\[0.1cm]

    \begin{subfigure}[h]{0.49\textwidth}
    \includegraphics[width=\textwidth]{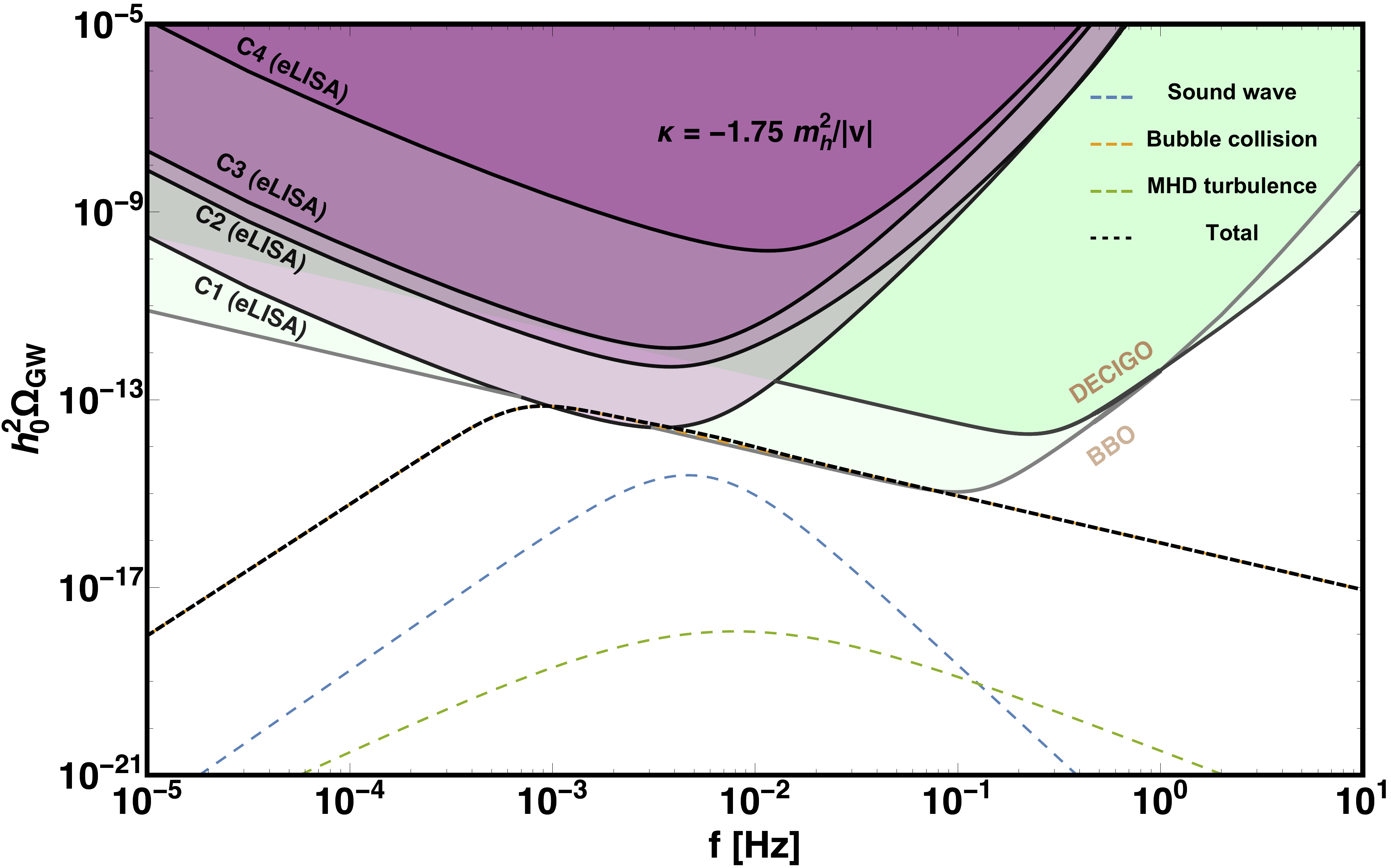}%
    \caption{}\label{fig:k175}
    \end{subfigure}
    \begin{subfigure}[h]{0.49\textwidth}
    \includegraphics[width=\textwidth]{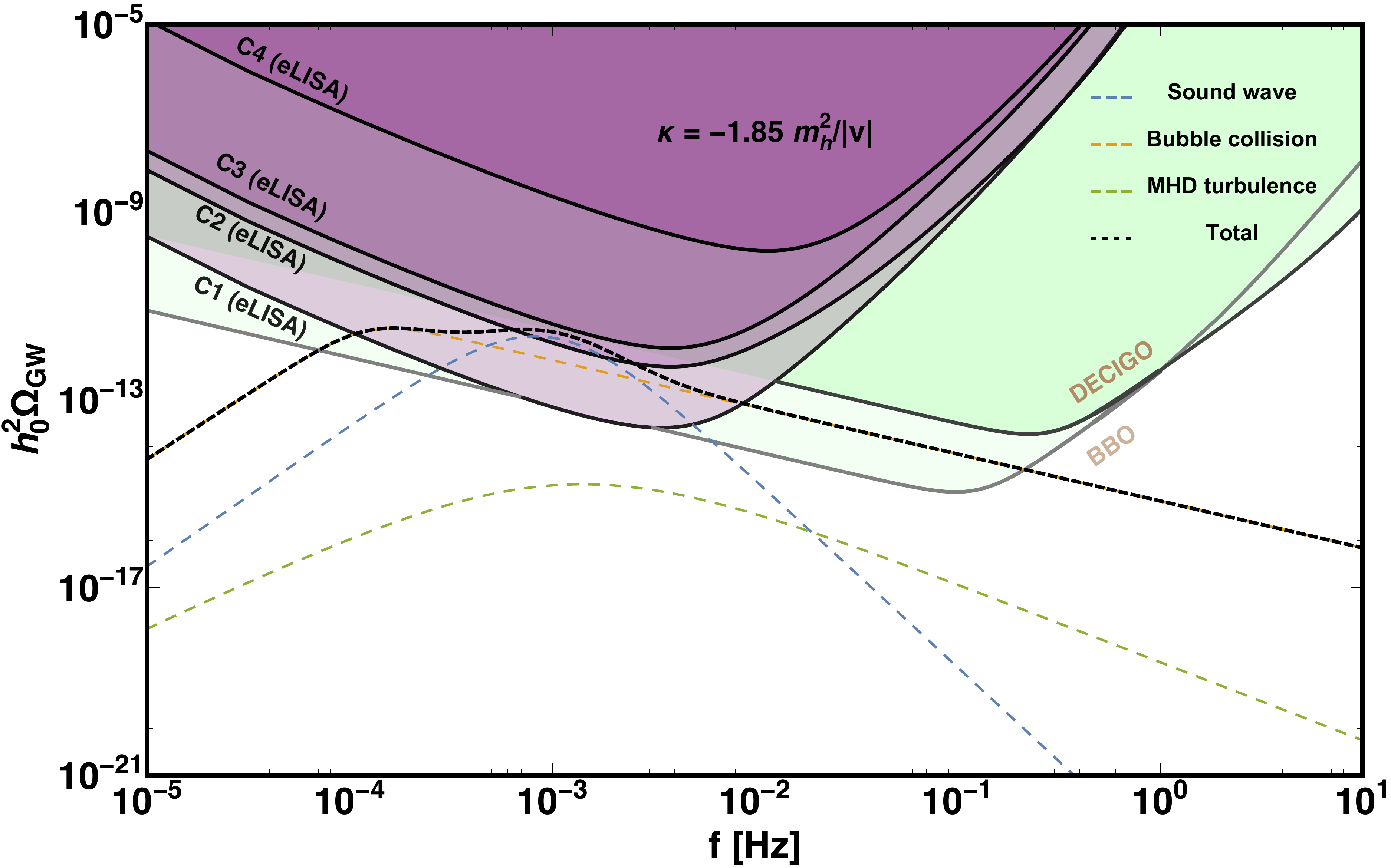}
    \caption{}\label{fig:k185}
    \end{subfigure}
    \caption{Gravitational wave spectral energy density for a range of $\kappa$ values showing the individual contributions from each production mechanism and an approximate total. The sensitivity curves for relevant space-based detectors are given by the shaded regions. The purple shaded regions denote the eLISA configurations C1 to C4 (see (\ref{app:eLISA})), the light green shaded region corresponds DECIGO and the grey curve to BBO. The cubic coupling $\kappa$  for each graph are as follows (\subref{fig:k125})  $-1.25 m_h^2/|v|$,  (\subref{fig:k15})  $-1.50 m_h^2/|v|$, (\subref{fig:k175}) $-1.75 m_h^2/|v|$ and (\subref{fig:k185}) $-1.85 m_h^2/|v|$. }
    \label{fig:results}
\end{figure}

In Fig.~\ref{fig:results}, the expected  gravitational wave spectra are shown  along with the sensitivity curves of future space based interferometers ---   evolved LISA (eLISA) \cite{AmaroSeoane:2012km}, Deci-Hertz Interferometer Gravitational-wave Observatory (DECIGO) \cite{Kawamura:2011zz}  and Big Bang observer (BBO) \cite{Corbin:2005ny,Harry:2006fi}. 
These detectors will be successors of the Laser Interferometer Space Antenna (LISA) project \cite{Danzmann:2003tv}, bridging its frequency reach to those that are ground based.  
The sensitivity for various configurations of  eLISA was adopted from   \cite{Caprini:2015zlo} (cf. App.~\ref{app:eLISA}), whereas the DECIGO and BBO sensitivity curves are from \cite{Yagi:2011wg}.

In these calculations, we have assumed a bubble wall velocity of $0.9$c, which is conservative for a run-away wall scenario and applicable to the analytical fit for sound waves, in equation (\ref{eq:sw}). 
Generically, the increase of $|\kappa|$ increases the strength of the transition (cf. Sec.~\ref{sec:fin_temp}) and the peak of the energy density spectra increase. This is also associate with a lowering of the characteristic frequencies. Since we are in the run-away regime, it is unsurprising that there the contribution from  bubble collisions gives the dominant source of gravitational waves. In agreement with \cite{Caprini:2015zlo,Hindmarsh:2015qta,Leitao:2012tx,Leitao:2015fmj}, we find little evidence from  magnetohydrodynamic sources, and that taking a less conservative efficiency of $\epsilon \sim 10\%$ will not change this conclusion. We find that gravitational waves can be observed for a narrow range of cubic coupling values $|\kappa|\in  [ 111,118] $ GeV. This will be dependent  on the type and configuration of the detector built.

%%%%%%%%%%%%%%%%%%%%%%%%%%%%%%%%%%%%%%%%%%%%%%
%%Section
%%%%%%%%%%%%%%%%%%%%%%%%%%%%%%%%%%%%%%%%%%%%%%
\section{Conclusions}\label{sec:conclusion}
In this paper we have studied the electroweak phase transition within the Standard Model with non-linearly realised electroweak gauge symmetry. Namely, we focused on the anomalous Higgs cubic coupling $\kappa$, which could drive strongly first order phase transition, if sufficiently large. With increase of $|\kappa|$, the nucleation  temperature, $T_n$, drops well below $W/Z$ masses, resulting in decreased  period of phase transition and in higher velocities of the nucleated bubbles. However for very large $|\kappa|$ values, the nucleation rate drops substantially and the universe is trapped in the high temperature phase. Thus, strongly first order phase transition is possible only for a limited range of the anomalous cubic coupling $|\kappa | \in [79,118]$ GeV. We have also found that for  $|\kappa|\in [111,118]$ GeV, gravitational waves in the  $0.1-10~\mu$Hz frequency range can be produced during the electroweak phase transition with a sizeable enough amplitude to be detectable by the planned eLISA. Recent results from LISA Pathfinder mission \cite{PhysRevLett.116.231101} are encouraging for the feasibility of the eLISA project, which, if implemented, can provide a complimentary information on the nature of electroweak symmetry and the cosmological  phase transition. This information will be particularly important since the measurement of the Higgs cubic coupling at high luminosity LHC is feasible only with $30-50$\% accuracy \cite{Lu:2015jza}.

%%%%%%%%%%%%%%%%%%%%%%%%%%%%%%%%%%%%%%%%%%%%%%
%%Acknowledgements
%%%%%%%%%%%%%%%%%%%%%%%%%%%%%%%%%%%%%%%%%%%%%%
\acknowledgments

We would like to thank Xavier Calmet, Shinya Kanemura and Matsui Toshiori for useful discussions. This work was partially supported by the Australian Research Council. AK was also supported in part by the Rustaveli National Science Foundation under the project No. DI/12/6-200/13.

%%%%%%%%%%%%%%%%%%%%%%%%%%%%%%%%%%%%%%%%%%%%%%
%%Appendix
%%%%%%%%%%%%%%%%%%%%%%%%%%%%%%%%%%%%%%%%%%%%%%
\appendix
%%%%%%%%%%%%%%%%%%%%%%%%%%%%%%%%%%%%%%%%%%%%%
%Appendix
%%%%%%%%%%%%%%%%%%%%%%%%%%%%%%%%%%%%%%%%%%%%%
\clearpage
\appendix
\section{Appendix}
%%%%%%%%%%%%%%%%%%%%%%%%%%%%%%%%%%%%%%%%%%%%%%
%Section
%%%%%%%%%%%%%%%%%%%%%%%%%%%%%%%%%%%%%%%%%%%%%%
\subsection{Field-Dependent Masses and Daisy-loop Corrections}\label{app:daisy}
The field-dependent masses included in the effective potential analysis of Sec.~\ref{sec:fin_temp} are given as follows:
\begin{equation}
\begin{aligned} 
\label{eq:dof_mass_1}
 n_h&= 1,  \hspace{20ex}    &m_h^2(\rho,T)   =&  3{\lambda}\rho^2+ 2\kappa\rho - \mu^2 \\
   %\quad  m_\chi^2 (\rho) ={\lambda}\rho^2 - \mu^2 \\
 n_Z &=  3,   &m_Z^2 (\rho)  =&  \frac{g^2_2+g^2_1}{4}\rho^2, \\
 n_W&=  6, &m_W^2 (\rho)   =&  \frac{g^2_2}{4}\rho^2, \\
 n_t&= -12,   &m_t^2 (\rho)  = &\frac{y_t^2}{2}\rho^2.  
\end{aligned}
\end{equation}
The inclusion of Daisy-diagrams can effectively be described by a shift in the respective boson masses (only longitudinal components) by their Debye correction (cf. e.g. \cite{Arnold:1992rz,Espinosa:1993bs}):
\begin{equation}
  \begin{aligned}
    m_h^2 &\rightarrow m_h^2 (\rho,T)&&=m_h^2(\rho)+\frac{1}{4} \lambda T^2+ \frac{1}{8}g_2^2 T^2 +  \frac{1}{16}(g_2^2+g^2_1) T^2+\frac{1}{4}y_t^2T^2,\\
    m_{W_L}^2 (\rho)&\rightarrow m_{W_L}^2(\rho,T) &&=m_{W}^2 (\rho)+\frac{11}{6}g_2^2T^2,\\
    m_{Z_L}^2 (\rho)&\rightarrow m_{Z_L}^2(\rho,T) &&= \frac{1}{2}\left[ m_Z^2(\rho) + \frac{11}{6}\left(g_2^2+g_1^2\right)T^2 + \Delta(\rho,T) \right],\\
    m_{\gamma_L}^2 (\rho)&\rightarrow m_{\gamma_L}^2(\rho,T)& &= \frac{1}{2}\left[ m_Z^2(\rho) + \frac{11}{6}\left(g_2^2+g_1^2\right)T^2 - \Delta(\rho,T) \right],\\
      \end{aligned}
\end{equation}
where:%{\color{red}[need verifiy from 1404.7673,1504.05949,9301285]}
\begin{equation}
  \Delta^2(\rho,T):=%m_Z^4(\rho) + \frac{11}{3}\frac{\cos^2 2\theta_W}{\cos^2 \theta_W}T^2\left[ m_Z^2(\rho)+ \frac{11}{12}\frac{g^2}{\cos^2 \theta_W}T^2 \right]
  \left( m_Z^2(\rho)+ \frac{11}{6}(g_2^2+g_1^2)T^2 \right)^2 -g_1^2g_2^2\frac{11}{3}T^2\left(\frac{11}{3}T^2+\rho^2  \right). 
\end{equation}
The number of degrees of freedom is then:
\begin{equation}
  g_{W_L}= 2g_{Z_L} = 2g_{\gamma_L}=2, \qquad g_{W_T}= 2g_{Z_T} =2 g_{\gamma_T}=4. 
\end{equation}

\subsection{eLISA Configurations}
\label{app:eLISA}

The proposed eLISA configurations used to estimate the sensitivity curves are tabulated below.

\begin{table}[h!]
%\fontsize{8pt}{9.6pt}\selectfont
\begin{center}
{\renewcommand{\arraystretch}{1.5}
\begin{tabular}{ c|c| c|c|c }
\hline 
%%%%%%%
Configuration Parameters & C1&C2&C3&C4\\\hline
Number of Arms & 3 & 3 & 2 & 2 \\\hline
Length per Arm ($10^9$m) & 5 & 1 & 2 & 1 \\\hline
Experiment Duration (Years) & 5 & 5 & 5 & 2 \\\hline

\end{tabular}}
\caption{eLISA configurations used in Fig.~\ref{fig:results}, as taken from \cite{Caprini:2015zlo}. 
  \label{tab:elisa_config}}
\end{center}
\end{table}

\bibliographystyle{mybibsty}
\bibliography{myrefs}

\end{document}